\documentclass[12pt,a4paper]{article}

\newcommand\non{\nonumber}

\def\disp{\displaystyle}

\newcommand{\Trace}{\mbox{\rm Tr}\,}

\newcommand{\uk}{\underline{k}}
\newcommand{\uj}{\underline{j}}
\newcommand{\C}{\widehat{\chi}}

\newtheorem{theorem}{Theorem}[section]
\newtheorem{lemma}{Lemma}[section]

\newtheorem{corollary}{Corollary}[section]

\def \outlineby #1#2#3{\vbox{\hrule\hbox{\vrule\kern #1%
\vbox{\kern #2 #3\kern #2}\kern #1\vrule}\hrule}}%
\def \endbox {\outlineby{4pt}{4pt}{}}%
\newcommand{\qed}{\hfill \endbox}
\newcommand{\be}{\begin{equation}}
\newcommand{\ee}{\end{equation}}
\newcommand{\bea}{\begin{eqnarray}}
\newcommand{\eea}{\end{eqnarray}}

\font\BB=msbm10
\newcommand{\NN}{\hbox{\BB N}}

\newcommand{\PP}{\hbox{\BB P}}
\newcommand{\EE}{\hbox{\BB E}}

\newcommand\third{\frac{1}{3}}
\newcommand\half{\frac{1}{2}}

\def\reff#1{(\ref{#1})}

\begin{document}

\begin{titlepage}

\begin{center}
{\Large{\bf{Coding Theorem for a Class of Quantum}}}
\end{center}

\begin{center}
{\Large{\bf{Channels
with Long-Term Memory}}}
\end{center}
\vspace{2cm}
\bigskip
\centerline{Nilanjana Datta} \centerline{Statistical Laboratory}
\centerline{Centre for Mathematical Sciences}
\centerline{University of Cambridge} \centerline{Wilberforce Road,
Cambridge CB30WB} \centerline{email: n.datta@statslab.cam.ac.uk}
\bigskip

\centerline{Tony C. Dorlas} \centerline{Dublin Institute for
Advanced Studies} \centerline{School of Theoretical Physics}
\centerline{10 Burlington Road, Dublin 4, Ireland.}
\centerline{email: dorlas@stp.dias.ie }
\bigskip

\vspace{1cm}

\noindent
\narrower{\bf Keywords:} quantum channels with long-term memory,
classical capacity, Feinstein's Fundamental Lemma, Helstrom's Theorem. 
\bigskip

\begin{abstract} 
In this paper we consider the transmission of classical
information through a class of quantum channels with long-term memory,
which are given by convex combinations of product channels.
Hence, the memory of such channels is given by a Markov chain which 
is aperiodic but not irreducible. We prove the coding theorem and 
weak converse for this class of channels.
The main techniques that we employ, are a quantum version of Feinstein's
Fundamental Lemma \cite{Feinstein, Khinchin} and a generalization of Helstrom's Theorem.
\cite{helstrom}.
\end{abstract}
\end{titlepage}

\newpage

\section{Introduction}
The biggest hurdle in the path of efficient information
transmission is the presence of noise, in both classical and quantum
channels. This noise causes a distortion of the information sent
through the channel. Error--correcting codes are used to overcome
this problem. Instead of transmitting the original
messages, they are encoded into codewords, which are then
sent through the channel. Information transmission is said to be
reliable if the probability of error, in decoding the output of the
channel, vanishes asymptotically in the number of uses of the channel
(see e.g. \cite{Cover} and
\cite{Nielsen}). The aim is to achieve reliable transmission,
whilst optimizing the rate, i.e., the ratio between the size of the
message and its correponding codeword. The optimal rate of reliable
transmission is referred to as the capacity of the the channel.

Shannon, in his Noisy Channel Coding Theorem \cite{shannon},
obtained an explicit expression for the channel capacity
of discrete, memoryless\footnote{
For such a channel, the noise affecting successive input states,
is assumed to be perfectly uncorrelated.},
classical channels. The first
rigorous proof of this fundamental theorem was provided by
Feinstein \cite{Feinstein}. He used a packing argument (see e.g.\cite{Khinchin}
) to find a
lower bound to the maximal number of codewords that can be sent
through the channel reliably, i.e., with an arbitrarily
low probability of error.
More precisely, he proved that for any given $\delta >0$, and sufficiently
large number, $n$, of uses of a memoryless classical channel, the lower bound
to the maximal number, ${N_n}$, of codewords that can be transmitted through the
channel reliably, is given by $${N_n} \ge 2^{n(H(X:Y)-\delta)}.$$ Here
$H(X:Y)$ is the mutual information of the random variables $X$ and $Y$,
corresponding to the input and the output of the channel, respectively.
We refer to this result as {\em{Feinstein's Fundamental Lemma}}, following
Khinchin \cite{Khinchin}. It implies that for a real number 
$R <C$, where $C = \max \, H(X:Y)$,
(the maximum being taken over all possible input distributions), 
$M_n \le 2^{nR}$
classical messages can be transmitted through the channel reliably. In
other words, any rate $R <C$ is {\em{achievable}}.

For real world communication channels, the {assumption} that 
noise is {uncorrelated} between {successive uses} of a channel cannot 
be {justified.} Hence {memory effects }need to be taken into account.
This leads us to the consideration of quantum channels with
memory. The {first model} of such a channel was studied by
{Macchiavello and Palma} \cite{chiara}. They showed that the
transmission of {classical information} through {two successive
uses} of a {quantum depolarising channel, with Markovian
correlated noise,} is enhanced by using inputs entangled over the
two uses. An important class of quantum channels with memory
consists of the so-called {\em{forgetful channels}}. The channel
studied in \cite{chiara} falls in this class. Roughly speaking, a
forgetful channel is one for which the output after a large number
of successive uses, does not depend on the initial input state.
Forgetful channels have been studied by Bowen and Mancini
\cite{bowen} and more recently by Kretschmann and Werner
\cite{KW}. In \cite{KW}, coding theorems for arbitrary forgetful
channels were proved. The proof of the direct channel coding
theorem for a class of quantum channels with Markovian correlated
noise, where the underlying Markov Chain was aperiodic and irreducible,
was sketched out in \cite{DD1}.
Very recently Bjelakovi\'c and Boche \cite{berlin} have proved a coding
theorem
for causal ergodic classical-quantum channels with decaying input memory.

The capacities of channels with long-term memory (i.e., channels
which are ``not forgetful''), had remained an open problem to
date. In this paper we evaluate the classical capacity of a class
of quantum channels with long-term memory. The tool that we
develop to prove the relevant coding theorem, can be considered to
be a quantum analogue of Feinstein's Fundamental Lemma \cite{DD1}. For a
quantum memoryless channel, our method yields an alternative proof
of the Holevo-Schumacher-Westmoreland (HSW) Theorem \cite{holevo,
SW}, similar in spirit to the proof in \cite{winter}.

We start the main body of our paper with some preliminaries in
Section \ref{prelim}.
Our main result is stated in Section \ref{mainresult}.
For clarity of exposition, we follow this with a proof of the quantum
analogue of Feinstein's Fundamental Lemma for memoryless channels
in Section \ref{qfein}.
The proof of our main result, for a class of quantum channels with long-term
memory, is given in Section \ref{nonerg}.

In summary, in this paper we consider the transmission of classical
information through a class of quantum channels with long-term memory,
which are convex combinations of product channels 
(defined through \reff{def_channel} of Section \ref{mainresult}).
The memory of the channel is given by a Markov chain which is aperiodic but not
irreducible. We prove the coding theorem 
and weak converse for this class of channels.
The main techniques that we employ are a quantum version of Feinstein's
Fundamental Lemma \cite{Feinstein, Khinchin} 
and a generalization of Helstrom's Theorem \cite{helstrom}.
Our results can be extended to quantum channels with arbitrary Markovian correlated noise. The proofs in this case are technically more involved and will
be presented in a subsequent paper.

\section{Preliminaries}
\label{prelim}
Let ${\cal B}({\cal H})$ denote the algebra of linear operators acting on
a finite--dimensional Hilbert space ${\cal H}$.
 The von Neumann entropy of a state $\rho$, i.e., a positive operator
of unit trace in ${\cal B}({\cal H})$, is defined as $S(\rho) =
-\Trace \rho \log \rho$, where the logarithm is taken to base $2$.
A quantum channel is given by a completely positive
trace--preserving (CPT) map $\Phi: {\cal B}({\cal H}) \to {\cal
B}({\cal K})$, where ${\cal H}$ and ${\cal K}$ are the input and
output Hilbert spaces of the channel. Let ${\hbox{dim }} {\cal
H}=d$ and ${\hbox{dim }} {\cal K}=d'.$ For any ensemble
$\{p_j,\rho_j\}$ of states $\rho_j$ and probability distributions
$\{p_j\}$, the Holevo $\chi$ quantity is defined as \be
\chi(\{p_j,\rho_j\}) := S\left( \sum_{j} p_j \,\rho_j
\right) - \sum_{j} p_j\, S(\rho_j). \label{Holevo} \ee The
Holevo capacity of a quantum channel $\Phi$ is given by
\begin{equation} \chi^*(\Phi) := \max_{\{p_j,\rho_j\}}\chi
\bigl(\{p_j,\Phi(\rho_j)\}\bigr),
\label{capacity} \end{equation} where the maximum is
taken over all ensembles $\{p_j,\rho_j\}$ of possible input states
$\rho_j \in {\cal B}({\cal H})$ occurring with probabilities $p_j$.
It is known that the maximum in (\ref{capacity}) can be achieved by
using an ensemble of pure states, and that it suffices to restrict the
maximum to ensembles of at most $d^2$ pure states.

Let us consider the transmission of classical information through
successive uses of a quantum channel $\Phi$. Let $N$ uses of the
channel be denoted by $\Phi^{(n)}$. Suppose Alice has a set of
messages, labelled by the elements of the set ${\cal{M}}_n = \{1,2,
\ldots, M_n\},$ which she would like to communicate to Bob, using
the quantum channel $\Phi$. To do this, she encodes each message
into a quantum state of a physical system with Hilbert space
${\cal{H}}^{\otimes n}$, which she then sends to Bob through
$n$ uses of the quantum channel. In order to infer the message
that Alice communicated to him, Bob makes a measurement (described
by POVM elements) on the state that he receives. The encoding and
decoding operations, employed to achieve reliable transmission of
information through the channel, together define a quantum error
correcting code (QECC). More precisely, a code ${\cal{C}}^{(n)}$
of size $N_n$ is given by a sequence $\{\rho_i^{(n)},
E_i^{(n)}\}_{i=1}^{N_n}$ where each $\rho_i^{(n)}$ is a state in
${\cal{B}}({\cal{H}}^{\otimes n})$ and each $E_i^{(n)}$ is a
positive operator acting in ${\cal{K}}^{\otimes n}$, such that
$\sum_{i=1}^{N_n} E_i^{(n)} \le {I}_n$. Here $I_n$ denotes
the identity operator in ${\cal{B}}({\cal{K}}^{\otimes n})$.
Defining $E_n^0 = I_n - \sum_{i=1}^{N_n} E_i^{(n)}$, yields a
resolution of identity in ${\cal{K}}^{\otimes n}$. Hence,
$\{E_i^{(n)}\}_{i=0}^{N_n}$ defines a POVM. An output $i\ge 1$
would lead to the inference that the state (or codeword)
$\rho_i^{(n)}$ was transmitted through the channel $\Phi^{(n)}$,
whereas the output $0$ is interpreted as a failure of any
inference. 
The average probability of error for the code ${\cal{C}}^{(n)}$ is given by
\begin{equation}
P_e({\cal{C}}^{(n)}):= \frac{1}{N_n} \sum_{i=1}^{N_n} \left(1 - {\Trace}\bigl(
\Phi^{(n)}(\rho_i^{(n)}) E_i^{(n)}\bigr)\right),
\label{codeerr}
\end{equation}
If there exists an
$N \in {\bf{N}}$ such that for all $n \ge N$, there exists a sequence
of codes $\{{\cal{C}}^{(n)}\}_{n=1}^\infty$, of sizes $N_n \ge 2^{nR}$,
for which $P_e({\cal{C}}^{(n)})\rightarrow 0$ as $n \rightarrow \infty$, then
$R$ is said to be {\em{achievable}} rate.
\medskip

The capacity of ${\Phi}$ is defined as
\begin{equation}
C(\Phi) := \sup R,
\end{equation}
where $R$ is an achievable rate. If the codewords $\rho_i^{(n)}$, $i=1,2, \ldots, N_n$,
are restricted to product states in ${\cal B}({\cal H}^{\otimes n})$,
the capacity $C(\Phi)$ is referred to as the {\em{product state capacity}}.

\section{Main Result}
\label{mainresult}
In this paper we study a class of channels with long-term
memory. For a channel $\Phi$ in this class,
$\Phi^{(n)} : {\cal B}({\cal H}^{\otimes n})
\to {\cal B}({\cal K}^{\otimes n})$ and the action of $\Phi^{(n)}$
on any state $\rho^{(n)} \in {\cal B}({\cal H}^{\otimes n})$
is given as follows:
\begin{equation} \Phi^{(n)}(\rho^{(n)}) = \sum_{i=1}^M
\gamma_i \Phi_i^{\otimes n}(\rho^{(n)}),
\label{def_channel}
\end{equation}
where
$\Phi_i: {\cal B}({\cal H}) \to {\cal B}({\cal K})$,
($i=1,\dots,M$) are CPT maps and $\gamma_i > 0$, $ \sum_{i=1}^M
\gamma_i = 1$. Notice that this is an example of a quantum
channel with memory given by a Markov chain, which is aperiodic but not
irreducible \cite{Norris}.

Our main result is given by the following theorem.
\begin{theorem}
\label{mainthm} The product state capacity of a channel $\Phi$,
with long-term memory, defined through \reff{def_channel}, is given
by
$$
C(\Phi)= \sup_{\{p_j,\rho_j\}} \left[ {\bigwedge}_{i=1}^M
\chi_i(\{p_j,\rho_j\}) \right],
$$
where $\chi_i(\{p_j,\rho_j\}):= \chi\left(\{p_j, \Phi_i(\rho_j)\}\right)$.
The supremum is taken
over all finite ensembles of states $\rho_j\in {\cal{B}}({\cal{H}})$ 
with probabilities
$p_j$.
\end{theorem} Here we use the standard notation $\bigwedge$ to
denote the minimum.

The product state capacity can be generalized to give the
classical capacity of the channel $\Phi$ in the usual manner, that
is, by considering inputs which are product states over uses of
blocks of $n$ channels, but which may be entangled across
different uses within the same block. The classical capacity
$C_{\hbox{\small{classical}}}(\Phi)$ is obtained in the limit $n
\rightarrow \infty$ and is given by \be
C_{\hbox{\small{classical}}}(\Phi) = \lim_{n \rightarrow \infty}
\frac{1}{n} C(\Phi^{(n)}). \ee

\section{Analogue of Feinstein's Fundamental Lemma for a Memoryless
Quantum Channel}
\label{qfein}

In this section we prove an analogue of Feinstein's Fundamental Lemma
 \cite{Feinstein}
for a memoryless quantum channel $\Phi$. This is given by Theorem
\ref{memoryless} below.
It provides an upper bound
to the maximal number of codewords that can be sent reliably through
$\Phi$.

The proof of our main result, Theorem \ref{mainthm}, employs a theorem
which is a generalization of Theorem \ref{memoryless}.

\begin{theorem} \label{memoryless}
Let $\Phi: {\cal B}({\cal H}) \to {\cal B}({\cal K})$
be a memoryless quantum channel.
Given $\epsilon > 0$, there exists $n_0 \in \NN$ such that for all $n\geq n_0$
there exist at least ${N_n} \geq 2^{n(\chi^*(\Phi)-\epsilon)}$ product
states
${\tilde \rho}^{(n)}_1, \dots, {\tilde \rho}^{(n)}_{{N_n}} \in {\cal
B}({\cal H}^{\otimes n})$ and positive operators $E_1^{(n)},\dots, E_{{N_n}}^{(n)} \in {\cal
B}({\cal K}^{\otimes n})$ such that $\sum_{k=1}^{{N_n}} E_k^{(n)} \leq I_n$ and
\begin{equation} \Trace \left[ \Phi^{\otimes n} \left( {\tilde
\rho}^{(n)}_k \right) E_k^{(n)} \right] > 1-\epsilon, \end{equation}
for each $k$.
\medskip

\noindent
Here $\chi^*(\Phi)$ is the Holevo capacity \reff{capacity}
of the memoryless quantum channel $\Phi$.
\end{theorem}

Before giving the proof of Theorem \ref{memoryless}, let us
briefly sketch the idea behind it. The proof employs the idea of
construction of a maximal code. For a given $\epsilon >0$,
starting with an empty code, the proof gives a prescription for
successively adding codewords $\rho_j^{(n)}$ and corresponding
POVM elements $E_j^{(n)}$, $j=1,2, \ldots,$ such that \be
\varepsilon_j^{(n)}:= 1 - \Trace \Phi^{(n)}(\rho_j^{(n)})\le
\epsilon \label{cond} \ee Note that $\varepsilon_j^{(n)}$ is the
probability of error in inferring the $j^{th}$ codeword. This is
done until no more codewords can be added without violating the
condition \reff{cond}. The resulting code is maximal. Let the size
of this code be ${N_n}$. The proof ensures that the number ${N_n}$
is large and provides a lower bound for it in terms of the Holevo
capacity $\chi^*(\Phi)$.

\textbf{Proof.} Let the maximum in (\ref{capacity}) be attained
for an ensemble $\{ p_j, \rho_j\}_{j=1}^J$. Denote $\sigma_j =
\Phi(\rho_j)$, ${\bar \sigma} = \sum_{j=1}^J p_j \Phi(\rho_j)$ and
${\bar \sigma}_n = {\bar\sigma}^{\otimes n}$. Since ${\bar \sigma}_n$
is a product state, its eigenvalues and eigenvectors can be labelled
by sequences $\uk=(k_1, \ldots, k_n) \in J^n$.

Choose $\delta > 0$. We will relate $\delta$ to $\epsilon$ at a later
stage. There exists $n_1 \in \NN$ such that for $n \geq n_1$,
there is a typical subspace ${\overline{\cal T}}_{\epsilon}^{(n)}$ of
${\cal{K}}^{\otimes n}$, with projection ${\bar P}_n$ such that if
${\bar\sigma}_n$ has a spectral decomposition \begin{equation}
{\bar \sigma}_n = \sum_{\uk} {\bar \lambda}_{\uk}^{(n)}
|\psi^{(n)}_{\uk} \rangle \langle \psi_{\uk}^{(n)}| \end{equation}
then
\begin{equation} \left| \frac{1}{n} \log {\bar
\lambda}_{\uk}^{(n)} + S({\bar \sigma}) \right| < \frac{\epsilon}{3}
\label{Tbar} \end{equation} for all $\uk$ such that
$|\psi_{\uk}^{(n)} \rangle \in \overline{{\cal T}}_{\epsilon}^{(n)}$
and
\begin{equation} \Trace ({\bar P}_n {\bar \sigma}_n) > 1-\delta^2.
\label{Pn} \end{equation}

Further define \begin{equation} {\bar S} = \sum_{j=1}^J p_j\, S(\sigma_j).
\label{Sbar} \end{equation}

\begin{lemma} Given a sequence $\uj = (j_1,\dots,j_n) \in J^n$, let
$P_{\uj}^{(n)}$ be the projection onto the subspace of ${\cal{K}}^{\otimes n}$
spanned by
the eigenvectors of $\sigma_{\uj}^{(n)} = \sigma_{j_1} \otimes
\dots \otimes \sigma_{j_n}$ with eigenvalues $ \lambda_{\uj,
\uk}^{(n)} = \prod_{i=1}^n \lambda_{j_i,k_i}$ such that
\begin{equation} \left| \frac{1}{n} \log \lambda_{\uj, \uk}^{(n)}
+ {\bar S} \right| < \frac{\epsilon}{3}. \end{equation} For any
$\delta > 0$ there exists $n_2 \in \NN$ such that for $n \geq
n_2$,
\begin{equation} \EE \left( \Trace \left( \sigma_{\uj}^{(n)}
P_{\uj}^{(n)} \right) \right) > 1-\delta^2. \end{equation}
\label{L1} \end{lemma}

\textbf{Proof.} Define i.i.d. random variables $X_1,\dots,X_n$
with distribution given by \begin{equation} \PP(X_i =
\lambda_{j,k}) = p_j\,\lambda_{j,k}, \end{equation} where
$\lambda_{j,k},\ k=1,2,\dots,d'$ are the eigenvalues of
$\sigma_j$. By the Weak Law of Large Numbers, \begin{eqnarray}
\frac{1}{n} \sum_{i=1}^n \log X_i \to \EE(\log X_i) &=&
\sum_{j=1}^J \sum_{k=1}^{d'} p_j\,\lambda_{j,k} \log \lambda_{j,k}
\non\\ &=& - \sum_{j=1}^J p_j\,S(\sigma_j) = -{\bar S}.
\end{eqnarray} It follows that there exists $n_2$ such that for $n
\geq n_2$, the typical set $T_{\epsilon}^{(n)}$ of sequences of pairs
$((j_1,k_1), \dots,(j_n,k_n))$ such that  \begin{equation} \left|
\frac{1}{n} \sum_{i=1}^n \log \lambda_{j_i,k_i} + {\bar S} \right|
< \frac{\epsilon}{3} \end{equation} satisfies \begin{equation} \PP
\left( T_{\epsilon}^{(n)} \right) = \sum_{((j_1,k_1),\dots, (j_n,k_n))
\in T_{\epsilon}^{(n)}} \prod_{i=1}^n p_{j_i} \lambda_{j_i,k_i}
> 1-\delta^2. \end{equation} Obviously,
\begin{equation} P_{\uj}^{(n)} \geq 
\sum_{\uk= (k_1, \ldots, k_n)\atop{((j_1,k_1),\dots, (j_n,k_n))
\in T_{\epsilon}^{(n)}}}
|\psi_{\uj,\uk}^{(n)} \rangle \langle
\psi_{\uj,\uk}^{(n)} | \end{equation} and \begin{equation}
\EE \left( \Trace \left( \sigma_{\uj}^{(n)} P_{\uj}^{(n)}
\right) \right) \geq \PP \left( T_{\epsilon}^{(n)} \right) >
1-\delta^2. \end{equation} \qed

Continuing the proof of the theorem, let ${{N_n}}$ be the maximal
number $N$ for which there exist product states ${\tilde
\rho}_1^{(n)}, \dots, {\tilde \rho}_N^{(n)}$ on ${\cal H}^{\otimes
n}$ and positive operators $E_1^{(n)},\dots, E_N^{(n)}$ on ${\cal
K}^{\otimes n}$ such that \begin{enumerate} \item[(i)] $
\sum_{k=1}^{N_n} E_k^{(n)} \leq {\bar P}_n$ and \item[(ii)]
$\Trace [\,{\tilde \sigma}_k^{(n)} E_k^{(n)} ] > 1-\epsilon$ and
\item[(iii)] $\Trace [\,{\bar \sigma}_n E_k^{(n)} ] \leq
2^{-n[S({\bar \sigma})-{\bar S}-\frac{2}{3} \epsilon]}$.
\end{enumerate}
Here ${\tilde \sigma}_k^{(n)} = \Phi^{\otimes n}({\tilde
\rho}_k^{(n)})$.

For any given $\uj \in J^n$ define
\begin{equation} V_{\uj}^{(n)} = \left( {\bar P}_n - \sum_{k=1}^{N_n}
E_k^{(n)} \right)^{1/2} {\bar P}_n P_{\uj}^{(n)} {\bar P}_n \left(
{\bar P}_n - \sum_{k=1}^{N_n} E_k^{(n)} \right)^{1/2}.
\end{equation} Clearly, $V_{\uj}^{(n)} \leq {\bar P}_n -
\sum_{k=1}^{N_n} E_k^{(n)}$, and we also have:

\begin{lemma} \begin{equation}
\Trace ({\bar \sigma}_n V_{\uj}^{(n)}) \leq 2^{-n[S({\bar
\sigma}) - {\bar S} - \frac{2}{3}\epsilon]}. \end{equation} \label{L2}
\end{lemma}

\textbf{Proof.} Put $Q_n = \sum_{k=1}^{N_n} E_k$. Note that $Q_n$
commutes with ${\bar P}_n$. Using the fact that ${\bar P}_n {\bar
\sigma}_n {\bar P}_n \leq 2^{-n [S({\bar \sigma})-
\frac{1}{3}\epsilon]}$ by (\ref{Tbar}), we have
\begin{eqnarray} \Trace ( {\bar \sigma}_n V_{\uj}^{(n)}) &=& \Trace
\left[ {\bar \sigma}_n ({\bar P}_n - Q_n)^{1/2} {\bar P}_n
P_{\uj}^{(n)} {\bar P}_n ({\bar P}_n - Q_n)^{1/2} \right] \non
\\ &=& \Trace \left[ {\bar P}_n {\bar \sigma}_n {\bar P}_n ({\bar P}_n - Q_n)^{1/2}
P_{\uj}^{(n)} ({\bar P}_n - Q_n)^{1/2} \right] \non \\ &\leq &
2^{-n[S({\bar \sigma})-\third\epsilon]} \Trace \left[ ({\bar P}_n -
Q_n)^{1/2} P_{\uj}^{(n)} ({\bar P}_n - Q_n)^{1/2} \right] \non \\
&\leq & 2^{-n[S({\bar \sigma})-\third\epsilon]} \Trace\,
(P_{\uj}^{(n)} ) \leq 2^{-n[S({\bar \sigma})- {\bar S} -
\frac{2}{3} \epsilon]}, \end{eqnarray} where, in the last inequality,
we used the standard upper bound on the dimension of the typical
subspace: $\Trace(P_{\uj}^{(n)}) \leq 2^{n[{\bar S}+\third
\epsilon]}$, which follows from Lemma~\ref{L1}. \qed

Since ${N_n}$ is maximal, it now follows that
\begin{equation} \Trace \left( \sigma_{\uj}^{(n)} V_{\uj}^{(n)}
\right) \leq 1-\epsilon. \label{Vj} \end{equation} and hence

\begin{corollary}
\begin{equation} \EE \left( \Trace \left[
\sigma_{\uj}^{(n)} V_{\uj}^{(n)} \right] \right) < 1-\epsilon.
\end{equation} \end{corollary}

\begin{lemma} For all $\eta > 0$, there exists $n_3 \in \NN$ such
that for all $n \geq n_3$, \begin{equation} \EE \left( \Trace
\left[ \sigma_{\uj}^{(n)} {\bar P}_n P_{\uj}^{(n)} {\bar
P}_n \right] \right) > 1-\eta. \end{equation} \label{L3}
\end{lemma}

\textbf{Proof.} We write \begin{eqnarray}
\EE \left(
\Trace \left[ \sigma_{\uj}^{(n)} {\bar P}_n P_{\uj}^{(n)}
{\bar P}_n \right] \right) &=&  \EE \left( \Trace
\left[ \sigma_{\uj}^{(n)} P_{\uj}^{(n)} \right] \right) -
\EE \left( \Trace \left[ \sigma_{\uj}^{(n)} (I_n - {\bar P}_n)
P_{\uj}^{(n)} \right] \right) \non \\ && \qquad - \EE \left(
\Trace \left[ \sigma_{\uj}^{(n)} {\bar P}_n P_{\uj}^{(n)}
(I_n - {\bar P}_n) \right] \right). \end{eqnarray} By
Lemma~\ref{L1}, the first term is $> 1-\delta^2$ provided $n \geq
n_2$. The last two terms can be bounded using the Cauchy-Schwarz
inequality as
follows:
\begin{eqnarray} && \EE \left( \Trace \left[
\sigma_{\uj}^{(n)} (I_n - {\bar P}_n) P_{\uj}^{(n)} \right]
\right)\nonumber\\ &&\,\,= \EE \left( \Trace \left[
\left(\sigma_{\uj}^{(n)} \right)^{1/2} (I_n - {\bar P}_n)
P_{\uj}^{(n)} \left(\sigma_{\uj}^{(n)} \right)^{1/2} \right]
\right) \non \\ &&\,\,\leq  \left\{ \EE \left( \Trace \left[
(I_n-{\bar P}_n) \sigma_{\uj}^{(n)} (I_n -{\bar P}_n) \right]
\right) \right\}^{1/2} \non \\ && \,\,\quad \times \left\{ \EE \left(
\Trace \left[ \left(\sigma_{\uj}^{(n)} \right)^{1/2}
P_{\uj}^{(n)} \left(\sigma_{\uj}^{(n)} \right)^{1/2} \right]
\right) \right\}^{1/2} \non \\ &&\,\,= \left\{ \EE \left( \Trace
\left[ \sigma_{\uj}^{(n)} (I_n -{\bar P}_n) \right] \right)
\right\}^{1/2} \left\{ \EE \left( \Trace \left[
\sigma_{\uj}^{(n)} P_{\uj}^{(n)} \right] \right)
\right\}^{1/2} \non \\ &&\,\,\leq \left\{ \EE \left( \Trace \left[
\sigma_{\uj}^{(n)} (I_n -{\bar P}_n) \right] \right)
\right\}^{1/2} \non \\ &&\,\,= \left( \Trace \left[ {\bar\sigma}_n
(I_n -{\bar P}_n) \right] \right)^{1/2} \leq \delta
\end{eqnarray} by (\ref{Pn}) provided $n \geq n_1$. Analogously,
\begin{equation} \EE \left( \Trace \left[
\sigma_{\uj}^{(n)} {\bar P}_n P_{\uj}^{(n)} (I_n - {\bar P}_n)
\right] \right)  \leq \delta.
\end{equation}
Choosing $n_3 = n_1 \vee n_2$ and $\delta^2 + 2 \delta < \eta$ the
result follows. \qed

\begin{lemma} Assume $\eta < \third \epsilon$. Then for $n \geq n_3$,
\begin{equation} \Trace
\left[ {\bar\sigma}_n \sum_{k=1}^N E_k^{(n)} \right] = \EE \left( \Trace
\left[ \sigma_{\uj}^{(n)} \sum_{k=1}^N E_k^{(n)} \right] \right) \geq
\eta^2. \end{equation}  \label{L4}
\end{lemma}

\textbf{Proof.} Define \begin{equation} Q'_n = {\bar P}_n - ({\bar
P}_n-Q_n)^{1/2}. \end{equation} By the above corollary,
\begin{eqnarray} 1-\epsilon &\geq& \EE \left\{ \Trace \left(
\sigma_{\uj}^{(n)} ({\bar P}_n-Q'_n) P_{\uj}^{(n)} ({\bar
P}_n-Q'_n) \right) \right\} \non \\ &=& \EE \left\{ \Trace \left(
\sigma_{\uj}^{(n)} {\bar P}_n P_{\uj}^{(n)} {\bar P}_n
\right) \right\} \non \\ && - \EE \left\{ \Trace \left(
\sigma_{\uj}^{(n)} Q'_n P_{\uj}^{(n)} {\bar P}_n \right)  +
\Trace \left( \sigma_{\uj}^{(n)} {\bar P}_n P_{\uj}^{(n)}
Q'_n \right) \right\} \non \\ && + \EE \left\{ \Trace \left(
\sigma_{\uj}^{(n)} Q'_n P_{\uj}^{(n)} Q'_n \right) \right\}.
\end{eqnarray} Since the last term is positive, we have, by
Lemma~\ref{L3},
\begin{equation} \EE \left\{ \Trace \left( \sigma_{\uj}^{(n)}
Q'_n P_{\uj}^{(n)} {\bar P}_n \right)  + \Trace \left(
\sigma_{\uj}^{(n)} {\bar P}_n P_{\uj}^{(n)} Q'_n \right)
\right\} \geq \epsilon - \eta > 2 \eta.
\end{equation}  On the other hand,
using Cauchy-Schwarz for each term, we have \begin{eqnarray}
\lefteqn{ \EE \left\{ \Trace \left( \sigma_{\uj}^{(n)} Q'_n
P_{\uj}^{(n)} {\bar P}_n \right)  + \Trace \left(
\sigma_{\uj}^{(n)} {\bar P}_n P_{\uj}^{(n)} Q'_n \right)
\right\}} \non \\ &\leq& 2 \left\{ \EE \left[ \Trace \left(
Q'_n \sigma_{\uj}^{(n)} Q'_n \right) \right] \right\}^{1/2}
\left\{ \EE \left[ \Trace \left( \sigma_{\uj}^{(n)} {\bar P}_n
P_{\uj}^{(n)} {\bar P}_n \right) \right] \right\}^{1/2} \non \\
&\leq & 2\left\{ \EE \left[ \Trace \left( \sigma_{\uj}^{(n)}
Q^{\prime 2}_n \right) \right] \right\}^{1/2}. \end{eqnarray}
Thus, \begin{equation}  \EE \left[ \Trace \left(
\sigma_{\uj}^{(n)} Q^{\prime 2}_n \right) \right] \geq \eta^2.
\end{equation} To complete the proof, we now claim that
\begin{equation} Q_n \geq (Q'_n)^2. \label{claim}\end{equation} Indeed,
on the domain of ${\bar P}_n$, \reff{claim}
follows  from the inequality
$1-(1-x)^2 \geq x^2$ for $0 \leq x \leq 1$. \qed

To complete the proof of Theorem \ref{memoryless}, we now have by assumption,
\begin{equation} \Trace \left[ {\bar \sigma}_n E_k^{(n)} \right] \leq
2^{-n[S({\bar \sigma}) - {\bar S} - \frac{2}{3} \epsilon]}
\end{equation} for all $k=1,\dots,{N_n}$. On the other hand, choosing
$\eta < \third \epsilon$ and $\delta < \third \eta$, we have by
Lemma~\ref{L4},
\begin{equation} \Trace
\left[ {\bar\sigma}_n \sum_{k=1}^{{N_n}} E_k^{(n)} \right] \geq \eta^2
\end{equation} provided $n \geq n_3$.
It follows that \begin{equation} {N_n} \geq \eta^2 2^{n [S({\bar
\sigma}) - {\bar S} - \frac{2}{3} \epsilon]} \geq 2^{n [S({\bar
\sigma}) - {\bar S} - \epsilon]} \end{equation} for $n \geq n_3$ and
$n \geq -\frac{6}{\epsilon} \log \eta.$ \qed

\section{A class of channels with long-term memory}
\label{nonerg}

We now consider the class of quantum channels with
long-term memory, mentioned in the Introduction:
\begin{equation} \Phi^{(n)}(\rho^{(n)}) = \sum_{i=1}^M
\gamma_i \Phi_i^{\otimes n}(\rho^{(n)}), \label{defch}\end{equation} where
$\Phi_i: {\cal B}({\cal H}) \to {\cal B}({\cal K})$,
($i=1,\dots,M$) are CPT maps and $\gamma_i > 0$, $ \sum_{i=1}^M
\gamma_i = 1$.

For an ensemble of states $\{p_j, \rho_j\}$ where $\rho_j \in
{\cal{B}}({\cal{H}})$, define \be \C(\{p_j,\rho_j\}) :=
{\bigwedge}_{i=1}^M \chi_i(\{p_j,\rho_j\}), \ee where
$\chi_i(\{p_j,\rho_j\})= \chi\left(\{p_j, \Phi_i(\rho_j)\} \right)$.

\subsection{Proof of the direct part of Theorem \ref{mainthm}}

To prove the direct part of Theorem \ref{mainthm}, i.e., the fact that
a rate $R< C(\Phi)$ is achievable, we employ
the quantum analogue of Feinstein's Fundamental Lemma for the class
of channels defined by \reff{defch}. This analogue is given by the following
theorem, which we prove in Section \ref{pffeinlong}
\begin{theorem} \label{fein_longterm}
Given $\epsilon >0$, there exists $n_0 \in \NN$ such that for all $n
\geq n_0$ there exist at least  $N_n \geq 2^{n(C(\Phi)-\epsilon)}$
product states $\rho^{(n)}_1, \dots, \rho^{(n)}_N \in {\cal
B}({\cal H}^{\otimes n})$ and positive operators $E_1^{(n)},\dots,
E_{N_n}^{(n)} \in {\cal B}({\cal K}^{\otimes n})$ such that
$\sum_{k=1}^{N_n} E_k^{(n)} \leq I_n$ and such that for each
$k=1,\dots,N_n$,
\begin{equation} \Trace \left[ \Phi^{(n)} \left( \rho^{(n)}_k
\right) E_k^{(n)} \right] > 1-\epsilon. \label{inderr}\end{equation}
Here \be C(\Phi) := \sup_{\{p_j,\rho_j\}} \left[
{\bigwedge}_{i=1}^M \chi_i(\{p_j,\rho_j\}) \right] =
\sup_{\{p_j,\rho_j\}} \C(\{p_j,\rho_j\}), \label{nonergcap} \ee
where  the supremum is over all finite ensembles of states
$\rho_j$ with probabilities $p_j$.
\end{theorem}

The above theorem implies that a rate $R <C(\Phi)$ is achievable.
This can be seen as follows: Given an $R <C(\Phi)$, choose $\epsilon >0$
such that $R<C(\Phi)- \epsilon.$ Then, Theorem \ref{fein_longterm} guarantees
the existence of codes ${\cal{C}}^{(n)}$ of size
$$N_n \ge 2^{n(C(\Phi) - \epsilon)} \ge 2^{nR},$$ with codewords given
by product states $\rho_j^{(n)}$, and POVM elements $E_j^{(n)}$,
for which the probability of error, $\varepsilon_j^{(n)}$, can be made arbitrarily small, for
each $j \in \{1,2, \ldots, N_n\}$ and
$n$ large enough. Hence the rate $R$ is acheivable.

\subsubsection{Proof of Theorem \ref{fein_longterm}}
\label{pffeinlong} Choose an ensemble $\{ p_j, \rho_j\}_{j=1}^J$
such that \begin{equation} C(\Phi) < \C(\{p_j,\rho_j\}) +
\frac{1}{4} \epsilon. \end{equation}  Define $\sigma_{i,j} =
\Phi_i(\rho_j)$, $\sigma_{i,\uj}^{(n)} = \otimes_{r=1}^n
\sigma_{i,j_r}$, ${\bar \sigma}_i = \sum_{j=1}^J p_j
\Phi_i(\rho_j) = \Phi_i({\bar \rho})$, and ${\bar \sigma}_i^{(n)}
= {\bar\sigma}_i^{\otimes n}$. Let ${\bar P}_i^{(n)}$,
$i=1,\dots,M$, be the orthogonal projections onto the typical
subspaces for the states ${\bar \sigma}_i^{(n)}$ so that, as
above,
\begin{equation} \Trace ({\bar P}_i^{(n)} {\bar \sigma}_i^{(n)}) >
1-\delta^2 \label{Pni}
\end{equation} for $n$ large enough, and \begin{equation}
{\bar P}_i^{(n)} {\bar \sigma}_i^{(n)} {\bar P}_i^{(n)} \leq 2^{-n
[S({\bar \sigma}_i)- \frac{1}{4}\epsilon]}. \end{equation} By
Lemma~\ref{L1} there also exist typical subspaces with projections
$P_{i,\uj}^{(n)}$ for which \begin{equation} \EE \left( \Trace
\left( \sigma_{i,\uj}^{(n)} P_{i,\uj}^{(n)} \right) \right)
> 1-\delta^2 \end{equation} for $n$ large enough.

To distinguish between the different memoryless branches,
$\Phi_i$, of the quantum channel $\Phi$, we add a preamble to the
input state encoding each message in the set ${\cal{M}}_n$. This is
given by an $m$-fold tensor product of a suitable state (as
described below). Let us first sketch the idea behind adding such
a preamble. Helstrom \cite{helstrom} showed that two states
$\sigma_1$ and $\sigma_2$, occurring with a priori probabilities
$\gamma_1$ and $\gamma_2$ respectively, can be distinguished, with
an asymptotically vanishing probability of error, if a suitable
collective measurement is performed on the $m$-fold tensor
products $\sigma_1^{\otimes m}$ and $\sigma_2^{\otimes m}$, for a
large enough $m \in \NN$. The optimal measurement is
projection-valued. The relevant projection operators, which we
denote by $\Pi^+$ and $\Pi^-$, are the orthogonal projections onto
the positive and negative eigenspaces of the difference operator
$A_m= \gamma_1\sigma_1^{\otimes n} - \gamma_2\sigma_2^{\otimes
n}$. Here we generalize this result to distinguish between the
different branches $\Phi_i$. If the preamble is given by a state
$\omega^{\otimes m}$, then, by using Helstrom's result, we can
construct a POVM which distinguishes between the output states
$\sigma_i^{\otimes n}:= \bigl(\Phi_i(\omega)\bigr)^{\otimes n}$
corresponding to the different branches $\Phi_i$, $i=1,2, \ldots,
M$. The outcome of this POVM measurement would in turn serve to
determine which branch of the channel is being used for
transmission.

Notice that we may assume that all branches $\Phi_i$ are
different. Indeed, otherwise we do not need to distinguish them
and can introduce a compound probability for each set of identical
branches. This assumption means that there exist states
$\omega_{i,j}$ on $\cal H$ for each pair $1 \leq i < j \leq M$
such that $\Phi_i(\omega_{i,j}) \neq \Phi_{j}(\omega_{i,j})$.
Introducing the fidelity of two states as in \cite{Nielsen},
\begin{equation} F(\sigma,\sigma') = \Trace \sqrt{\sigma^{1/2}
\sigma'\, \sigma^{1/2}}, \end{equation} we then have
\begin{equation} F(\Phi_i(\omega_{i,j}),\Phi_{j}(\omega_{i,j}))
\leq f < 1 \label{fidelitybnd} \end{equation} for all pairs
$(i,j)$.  We now introduce, for any $m \in \NN$ and $1 \leq i < j
\leq M$, the difference operators
\begin{equation} A^{(m)}_{i,j} = \gamma_i
\bigl(\Phi_i(\omega_{i,j})\bigr)^{\otimes m} - \gamma_{j}
\bigl(\Phi_{j}(\omega_{i,j})\bigr)^{\otimes m}. \end{equation}
Let $\Pi_{i,j}^\pm$ be the orthogonal projections onto the eigenspaces of
$A_{i,j}^{(m)}$ corresponding to all non-negative, and all
negative eigenvalues, respectively.

\begin{lemma} \label{LPi} Suppose that for a given $\delta >0$,
\begin{equation} |\Trace
[|A_{i,j}^{(m)}|] - (\gamma_i+\gamma_{j})| \leq \delta.
\end{equation} Then \begin{equation}
|\Trace [\Pi_{i,j}^+ \bigl(\Phi_i(\omega_{i,j})\bigr)^{\otimes m}] - 1| \leq
\frac{\delta}{2 \gamma_i}
\end{equation} and
\begin{equation}
|\Trace [\Pi_{i,j}^- \bigl(\Phi_{j}(\omega_{i,j})\bigr)^{\otimes m}]-1|
\leq \frac{\delta}{2 \gamma_{j}}. \end{equation}
\end{lemma}

\textbf{Proof.} Write $A = A_{i,j}^{(m)}$ and $\Pi^\pm =
\Pi_{i,j}^\pm$. First note that
\begin{eqnarray} \Trace\,[\Pi^\pm \,A] &=& \half \Trace \,[A \pm
(\Pi^+ - \Pi^-)A] \non \\ &=& \half \left( \Trace [A] \pm \Trace
[|A|] \right) \non \\ &=& \half (\gamma_i - \gamma_{j}) \pm \half
\Trace[|A|]
\end{eqnarray} so that we have by the assumption \begin{equation}
|\Trace\,[\Pi^+ \,A] - \gamma_i| \leq \half \delta \end{equation}
and \begin{equation} |\Trace\,[\Pi^- \,A] + \gamma_{j}| \leq \half
\delta. \end{equation} Now, writing $\sigma_i =
\bigl(\Phi_i(\omega_{i,j})\bigr)^{\otimes m}$ and $\sigma_{j} =
\bigl(\Phi_{j}(\omega_{i,j})\bigr)^{\otimes m}$ we have obviously,
$\Trace\,[\Pi^- \sigma_i] \geq 0$, and on the other hand,
\begin{equation} \gamma_i \Trace\,[\Pi^- \sigma_i] =
\Trace\,[\Pi^-\,A] + \gamma_{j} \Trace \,[\Pi^- \sigma_{j}] \leq
-\gamma_{j} + \half \delta + \gamma_{j} = \half \delta.
\end{equation} The first result thus follows from $\Pi^+ + \Pi^- =
I_m$ and $\Trace \sigma_i = 1.$ Similarly, \begin{equation}
\gamma_{j} \Trace\,[\Pi^+ \sigma_{j}] = -\Trace\,[\Pi^+\,A] +
\gamma_{i} \Trace \,[\Pi^+ \sigma_{i}] \leq -\gamma_{i} + \half
\delta + \gamma_{i} = \half \delta.
\end{equation}
\qed

To compare the outputs of all the different branches of the channel,
we define projections ${\tilde \Pi}_i$ on
the tensor product space $\bigotimes_{1 \leq i < j \leq M} {\cal
K}^{\otimes m} = {\cal K}^{\otimes mL}$ with $L = {M \choose 2}$
as follows: \begin{equation} {\tilde \Pi}_i = \bigotimes_{1 \leq i_1 <
i_2 \leq M} \Gamma_{i_1,i_2}^{(i)}, \mbox{ where } \Gamma_{i_1,i_2}^{(i)}
= \left\{ \begin{array}{lcl} I_m &\mbox{ if } &i_1 \neq i \mbox{
and } i_2 \neq i \\ \Pi_{i_1,i}^- &\mbox{ if } &i_2 = i \\
\Pi_{i,i_2}^+ &\mbox{ if } &i_1 = i. \end{array} \right.
\end{equation}
Notice that it follows from the fact that $\Pi_{i,j}^+
\Pi_{i,j}^- = 0$, that the projections ${\tilde \Pi}_i$ are also
disjoint: \begin{equation} {\tilde \Pi}_i {\tilde \Pi}_{j} = 0 \quad
{\hbox{for }} \, i\ne j.
\end{equation}

Introducing the notation \begin{equation} \omega^{(mL)} =
\bigotimes_{i_1 < i_2} \omega_{i_1,i_2}^{\otimes m},
\label{preamble}
\end{equation} we now have

\begin{lemma} \label{Ldistinct}
For all $i=1,\dots,M$, \begin{equation} \lim_{m \to \infty} \Trace
\left[ {\tilde \Pi}_i\, \Phi_i^{\otimes mL} \left( \omega^{(mL)}
\right) \right] = 1. \end{equation} \end{lemma}

\textbf{Proof.} Notice that for all $i < j$, \begin{equation}
F(\gamma_i \Phi_i(\omega_{i,j})^{\otimes m}, \gamma_{j}
\Phi_{j}(\omega_{i,j})^{\otimes m}) = \sqrt{\gamma_i
\gamma_{j}} F(\Phi_i(\omega_{i,j}), \Phi_i(\omega_{i,j}))^m <
f^m. \end{equation}  Using the inequalities \cite{Nielsen}
\begin{equation} \Trace (A_1) +
\Trace (A_2) - 2 F(A_1,A_2) \leq || A_1 - A_2 ||_1 \leq \Trace
(A_1) + \Trace (A_2)
\end{equation} for any two positive operators $A_1$ and $A_2$, we
find that
\begin{equation} \big|\,\Trace\,\bigl(|A_{i,j}^{(m)}|\bigr) -
(\gamma_i + \gamma_{j}) \big| \leq 2 f^m,
\end{equation}
since
\begin{equation} \Trace\,\bigl(|A_{i,j}^{(m)}|\bigr) = ||
\gamma_i \Phi_i(\omega_{i,j})^{\otimes m} - \gamma_{j}
\Phi_{j}(\omega_{i,j})^{\otimes m} ||_1. \end{equation}
Using Lemma~\ref{LPi} we then have
\begin{eqnarray} 1 \geq \lefteqn{\Trace \left[ {\tilde \Pi}_i
\Phi_i^{\otimes mL} \left( \bigotimes_{i_1 < i_2}
\omega_{i_1,i_2}^{\otimes m} \right) \right] =} \non \\ &=&
\prod_{i_1 < i} \Trace \left[ \Pi_{i_1,i}^-
\bigl(\Phi_i(\omega_{i_1,i})\bigr)^{\otimes m} \right] \,
\prod_{i_2 > i} \Trace \left[ \Pi_{i,i_2}^+
\bigl(\Phi_i(\omega_{i,i_2})\bigr)^{\otimes m} \right] \non \\
&\geq& \left(1-\frac{f^m}{\gamma_i}\right)^{M-1}. \end{eqnarray} \\ \qed

We now fix $m$ so large that \begin{equation} \Trace \left[
{\tilde \Pi}_i \,\Phi_i^{\otimes mL} \left( \omega^{(mL)} \right)
\right] > 1-\delta \label{Pitilde} \end{equation} for all
$i=1,\dots,M$. The product state $ \omega^{(mL)}$, defined through
\reff{preamble} is used as a preamble to the input state encoding
each message, and serves to distinguish between the different
branches, $\Phi_i$, $i=1,2,\ldots, M$, of the channel. If
$\rho_k^{(n)} \in {\cal{B}}({\cal{H}}^{\otimes n})$ is a product
state encoding the $k^{th}$ classical message in the set
${\cal{M}}_n$, then the $k^{th}$ codeword is given by the product
state $$ \omega^{(mL)} \otimes \rho_k^{(n)} .$$

\vskip1cm

Continuing with the proof of Theorem \ref{fein_longterm}, 
let $N={\tilde N}(n)$ be
the maximal number of product states ${\tilde \rho}_1^{(n)},
\dots, {\tilde \rho}_N^{(n)}$ on ${\cal H}^{\otimes n}$ (each of
which is a tensor product of states in the maximising ensemble
$\{p_j, \rho_j\}_{j=1}^J$) for which there exist positive
operators $E_1^{(n)}, \dots, E_N^{(n)}$ on ${\cal K}^{\otimes mL}
\otimes {\cal K}^{\otimes n}$ such that
\begin{enumerate}
\item[(i)] $  E_k^{(n)} = \sum_{i=1}^M {\tilde \Pi}_i \otimes
E_{k,i}^{(n)} $ and $ \sum_{k=1}^N E_{k,i}^{(n)} \leq {\bar
P}_i^{(n)}$ and \item[(ii)] $ \disp{ \sum_{i=1}^M \gamma_i \Trace
\left[\,{\tilde \Pi}_i \Phi_i^{\otimes mL}\left( \omega^{(mL)}
\right) \right] \Trace \left[\, \Phi_i^{\otimes n} \left(
{\tilde\rho}_k^{(n)} \right) E_{k,i}^{(n)} \right]
> 1-\epsilon} $ and
\item[(iii)] $ \disp{ \sum_{i=1}^M \gamma_i
\Trace \left[\,{\tilde \Pi}_i \Phi_i^{\otimes mL}\left(
\omega^{(mL)} \right) \right] \Trace \left[\, \left(\Phi_i \bigl( {\bar
\rho}\bigr)\right)^{\otimes n} E_{k,i}^{(n)} \right] \leq 2^{-n[C(\Phi)
- \half \epsilon]} }$.
\end{enumerate}
for ${\bar \rho}= \sum_{j=1}^J p_j \rho_j$.
For each $i=1,\dots,M$ and $\uj = (j_1,\dots,j_n) \in J^n,$ we define,
as before \begin{equation} V_{i,\uj}^{(n)} = \left( {\bar
P}_i^{(n)} - \sum_{k=1}^N E_{k,i}^{(n)} \right)^{1/2} {\bar
P}_i^{(n)} P_{i,\uj}^{(n)} {\bar P}_i^{(n)} \left( {\bar
P}_i^{(n)} - \sum_{k=1}^N E_{k,i}^{(n)} \right)^{1/2}.
\end{equation}
Clearly $ V_{i,\uj}^{(n)} \le {\bar P}_i^{(n)} - \sum_{k=1}^N E_{k,i}^{(n)}$.
Put \begin{equation} V_{\uj}^{(n)} :=
\sum_{i=1}^M {\tilde \Pi}_i \otimes V_{i,\uj}^{(n)}.
\end{equation}
This is a candidate for an additional measurement operator, $E_{N+1}^{(n)}$,
for Bob with corresponding input state $ {\tilde \rho}_{N+1}^{(n)} =
\rho_{\uj}^{(n)}= \rho_{j_1} \otimes \rho_{j_2} \ldots \otimes \rho_{j_n}$.
Clearly, the condition (i), given above, is satisfied and we also have

\begin{lemma} \label{L3.3}
\begin{equation} \sum_{i=1}^M \gamma_i
\Trace\,[ {\tilde \Pi}_i\,\Phi_i^{\otimes mL} \left(\omega^{(mL)}
\right)] \,\Trace \,[ {\bar\sigma}_i^{(n)} V_{i,\uj}^{(n)} ]
\leq 2^{-n[C(\Phi) - \half \epsilon]},
\end{equation}
where ${\bar\sigma}_i^{(n)} = \bigl(\Phi_i({\bar \rho})\bigr)^{\otimes n}$.
\end{lemma}

\textbf{Proof.} By Lemma~\ref{L2}, replacing $\third \epsilon$ by
$\frac{1}{4} \epsilon$ in the definition of the typical subspaces, we
have,
\begin{equation} \Trace ({\bar \sigma}_i^{(n)} V_{i,\uj}^{(n)})
\leq 2^{- n[S({\bar \sigma}_i) - {\bar S}_i - \half \epsilon]}
= 2^{- n[\chi_i -\half \epsilon]}.
\end{equation} for $n$ large enough. Then 
\begin{eqnarray} \sum_{i=1}^M \gamma_i
\Trace\,[ {\tilde \Pi}_i\,\Phi^{\otimes mL} \left(\omega^{(mL)}
\right)] \,\Trace \,[ {\bar\sigma}_i^{(n)} V_{i,\uj}^{(n)} ]
&\leq & \sum_{i=1}^M \gamma_i \Trace\,[ {\bar\sigma}_i^{(n)}
V_{i,\uj}^{(n)} ] \non \\ &\leq & \sum_{i=1}^M \gamma_i\, 2^{-
n[S({\bar \sigma}_i) - {\bar S}_i - \half \epsilon]} \non \\
&\leq & 2^{-n[\C(\Phi) - \half \epsilon]},\non \\
&\leq &  2^{-n[C(\Phi) - \frac{3}{4} \epsilon]},\non \\
\end{eqnarray} where we used
the obvious fact that $\Trace\,[ {\tilde \Pi}_i\,\Phi_i^{\otimes
mL}( \omega_i^{(mL)})] \leq 1$. \qed

By maximality of ${N}$ it now follows that the condition
(ii) above cannot hold,
that is, \begin{equation} \sum_{i=1}^M \gamma_i \Trace
\left[\,{\tilde \Pi}_i \Phi_i^{\otimes mL}\left( \omega^{(mL)}
\right) \right] \Trace \left[\, \Phi_i^{\otimes n} \left(
\rho_{\uj}^{(n)} \right) V_{i,\uj}^{(n)} \right] \leq 1-\epsilon
\end{equation} for every $\uj$, and this yields the following:
\begin{corollary}
\label{corr1}\begin{equation} \sum_{i=1}^M \gamma_i \Trace
\left[\,{\tilde \Pi}_i\, \Phi_i^{\otimes mL} \left( \omega^{(mL)}
\right) \right] \EE \left( \Trace \left[\, \Phi_i^{\otimes n}
\left( \rho_{\uj}^{(n)} \right) V_{i,\uj}^{(n)} \right]
\right) \leq 1-\epsilon. \end{equation}
\end{corollary}

\noindent

We also need the following lemma:

\begin{lemma} \label{L3.4} For all $\eta' > \delta^2 + 3 \delta$,
\begin{equation} \sum_{i=1}^M \gamma_i\,\Trace \left[\,
{\tilde \Pi}_i\, \Phi_i^{\otimes mL} \left( \omega^{(mL)} \right)
\right]\,\Trace \left[\,\sigma_{i,\uj}^{(n)} {\bar P}_i^{(n)}
P_{i,\uj}^{(n)} {\bar P}_i^{(n)} \right] > 1-\eta'
\end{equation} if $n$ is large enough. \end{lemma}

\textbf{Proof.} Using Lemma~\ref{L3} and (\ref{Pitilde}), we have
\begin{equation} \sum_{i=1}^M \gamma_i\, \Trace \left[\,{\tilde
\Pi}_i\, \Phi_i^{\otimes mL} \left( \omega^{(mL)} \right)
\right]\,\EE\left( \Trace \left[\,\sigma_{i,\uj}^{(n)} {\bar
P}_i^{(n)} P_{i,\uj}^{(n)} {\bar P}_i^{(n)} \right] \right) >
(1-\delta)(1-\eta)
\end{equation} provided $\eta>\delta^2 + 2\delta$.
Hence the result follows. \qed

\begin{lemma} Assume $\eta' < \frac{1}{3} \epsilon$ and write
\begin{equation} Q_i^{(n)} = \sum_{k=1}^N E_{k,i}^{(n)}.
\end{equation}
Then for $n$ large enough,
\begin{equation} \sum_{i=1}^M
\gamma_i \Trace \left[\,{\tilde \Pi}_i\, \Phi_i^{\otimes mL}
\left( \omega^{(mL)} \right) \right]\,\EE \left( \Trace \left[
\,\Phi_i^{(n)} \left( \rho_{\uj}^{(n)} \right) Q_i^{(n)}
\right] \right) \geq \eta^{\prime 2}. \end{equation}  \label{L3.5}
\end{lemma}

\textbf{Proof.} This is analogous to Lemma~\ref{L4}. Define
\begin{equation} Q_i^{(n)\prime} = {\bar P}_i^{(n)} -
({\bar P}_i^{(n)} - Q_i^{(n)})^{1/2}.
\end{equation} By the Corollary \ref{corr1},
\begin{eqnarray} 1- \epsilon &\geq& \sum_{i=1}^M \gamma_i \Trace
\left[\,{\tilde \Pi}_i\, \Phi_i^{\otimes mL} \left( \omega^{(mL)}
\right) \right] \EE \left( \Trace \left[\, \Phi_i^{\otimes n}
\left( \rho_{\uj}^{(n)} \right) V_{i,\uj}^{(n)} \right]
\right) \non \\ &=& \sum_{i=1}^M \gamma_i \Trace \left[\,{\tilde
\Pi}_i\, \Phi_i^{\otimes mL} \left( \omega^{(mL)} \right)
\right]\,\EE \left\{ \Trace \left( \sigma_{i,\uj}^{(n)} {\bar
P}_i^{(n)} P_{i,\uj}^{(n)} {\bar P}_i^{(n)} \right) \right\} \non \\
&& - \sum_{i=1}^M \gamma_i \Trace \left[\,{\tilde \Pi}_i\,
\Phi_i^{\otimes mL} \left( \omega^{(mL)} \right) \right] \,\EE
\left\{ \Trace \left( \sigma_{i,\uj}^{(n)} Q_i^{(n)\prime}
P_{i,\uj}^{(n)} {\bar P}_i^{(n)} \right) \right. \non \\ &&
\qquad \qquad \qquad \qquad \left. + \Trace \left(
\sigma_{i,\uj}^{(n)} {\bar
P}_i^{(n)} P_{i,\uj}^{(n)} Q_i^{(n)\prime} \right) \right\} \non \\
&& + \sum_{i=1}^M \gamma_i \Trace \left[\,{\tilde \Pi}_i\,
\Phi_i^{\otimes mL} \left( \omega^{(mL)} \right) \right]\,\EE
\left\{ \Trace \left( \sigma_{i,\uj}^{(n)} Q_i^{(n)\prime}
P_{i,\uj}^{(n)} Q_i^{(n)\prime} \right) \right\}. \non \\
\end{eqnarray} Since the last term is positive, we have, by
Lemma~\ref{L3.4},
\begin{eqnarray} && \sum_{i=1}^M \gamma_i \Trace \left[\,{\tilde \Pi}_i\,
\Phi_i^{\otimes mL} \left( \omega^{(mL)} \right) \right] \,\EE
\left\{ \Trace \left( \sigma_{i,\uj}^{(n)} Q_i^{(n) \prime}
P_{i,\uj}^{(n)} {\bar P}_i^{(n)} \right) \right. \non \\ &&
\qquad \qquad  \left. + \Trace \left( \sigma_{i,\uj}^{(n)}
{\bar P}_i^{(n)} P_{i,\uj}^{(n)} Q_i^{(n) \prime} \right)
\right\} \geq \epsilon - \eta' > 2 \eta'.
\end{eqnarray}  On the other hand,
using the Cauchy-Schwarz inequality for each term, we have \begin{eqnarray} &&
\sum_{i=1}^M \gamma_i \Trace \left[\,{\tilde \Pi}_i\,
\Phi_i^{\otimes mL} \left( \omega^{(mL)} \right) \right] \,\EE
\left\{ \Trace \left( \sigma_{i,\uj}^{(n)} Q_i^{(n)\prime}
P_{i,\uj}^{(n)} {\bar P}_i^{(n)} \right) \right. \non \\ &&
\qquad \qquad \qquad \qquad \left. + \Trace \left(
\sigma_{i,\uj}^{(n)} {\bar P}_i^{(n)} P_{i,\uj}^{(n)}
Q_i^{(n)\prime} \right) \right\} \leq  \non \\ && \quad \leq 2
\left\{ \sum_{i=1}^M \gamma_i \Trace \left[\,{\tilde \Pi}_i\,
\Phi_i^{\otimes mL} \left( \omega^{(mL)} \right) \right]\, \EE
\left[ \Trace \left( \sigma_{i,\uj}^{(n)}
(Q_i^{(n)\prime})^2 \right) \right] \right\}^{1/2} \non \\
&& \quad \times \left\{ \sum_{i=1}^M \gamma_i \Trace
\left[\,{\tilde \Pi}_i\, \Phi_i^{\otimes mL} \left( \omega^{(mL)}
\right) \right]\,\EE \left[ \Trace \left( \sigma_{i,\uj}^{(n)}
{\bar P}_i^{(n)} P_{i,\uj}^{(n)} {\bar P}_i^{(n)} \right)
\right] \right\}^{1/2} \non \\
&& \quad \leq 2\left\{\sum_{i=1}^M \gamma_i \Trace \left[\,{\tilde
\Pi}_i\, \Phi_i^{\otimes mL} \left( \omega^{(mL)} \right)
\right]\, \EE \left[ \Trace \left( \sigma_{i,\uj}^{(n)}
(Q_i^{(n)\prime})^2 \right) \right] \right\}^{1/2}. \end{eqnarray}
Thus, \begin{equation}  \sum_{i=1}^M \gamma_i \Trace
\left[\,{\tilde \Pi}_i\, \Phi_i^{\otimes mL} \left( \omega^{(mL)}
\right) \right]\, \EE \left[ \Trace \left( \sigma_{i,\uj}^{(n)}
(Q_i^{(n)\prime})^2 \right) \right] \geq \eta^{\prime\, 2}.
\end{equation} To complete the proof, we remark as before that
\begin{equation} Q_n \geq (Q'_n)^2. \end{equation} \qed

It now follows, as before, that for $n$ large enough, ${{\tilde
N}(n)} \geq (\eta')^2 \,2^{n[C(\Phi)- \frac{3}{4} \epsilon]}.$ We
take the following states as codewords:
\begin{equation} \rho_k^{(mL+n)} = \omega^{(mL)} \otimes
{\tilde \rho}_k^{(n)}. \end{equation} For $n$ sufficiently large
we then have \begin{equation} {N_{n+mL}} = {\tilde N}(n) \geq
(\eta')^2\,2^{n[C(\Phi)- \frac{3}{4} \epsilon]} \geq
2^{(mL+n)[C(\Phi)- \epsilon]}.
\end{equation}

To complete the proof we need to show that the set $E_k^{(n)}$
satisfies (\ref{inderr}). But this follows immediately from
condition (ii):
\begin{eqnarray} \lefteqn{\Trace \left[ \Phi^{(mL+n)}
\left( \rho^{(mL+n)}_k \right) E_k^{(n)} \right] =} \non \\ &=&
\sum_{i=1}^M \gamma_i \Trace \left[\,\Phi_i^{\otimes (mL+n)}
\left( \omega^{(mL)} \otimes
{\tilde \rho}^{(n)}_k \right) E_k^{(n)} \right] \non \\
&=& \sum_{i,j=1}^M \gamma_i \Trace \left[\,{\tilde \Pi}_{j}\,
\Phi_i^{\otimes mL} \left( \omega^{(mL)} \right) \right]\,\Trace
\left[ \Phi_i^{\otimes n} ( {\tilde \rho}^{(n)}_k) E_{k,j}^{(n)}
\right] \non \\ &\geq & \sum_{i=1}^M \gamma_i \Trace
\left[\,{\tilde \Pi}_{i}\, \Phi_i^{\otimes mL} \left(
\omega^{(mL)} \right) \right]\,\Trace \left[ \Phi_i^{\otimes n} (
{\tilde \rho}^{(n)}_k) E_{k,i}^{(n)} \right] > 1-\epsilon.
\end{eqnarray}

\qed

\subsection{{Proof of the converse of Theorem \ref{mainthm}}}

In this section we prove that it is impossible for Alice to
transmit classical messages reliably to Bob through the channel
$\Phi$ defined in \reff{defch} at a rate $R > C(\Phi)$.  This is
the weak converse of Theorem~\ref{mainthm} in the sense that the
probability of error 
does not tend to zero asymptotically as the length of
the code increases, for any code with rate $R > C(\Phi)$. To prove the
weak converse, suppose that Alice encodes messages labelled by
$\alpha \in {\cal M}_n$ by product states $\rho_\alpha^{(n)} =
\rho_{\alpha,1} \otimes \dots \otimes \rho_{\alpha,n}$ in ${\cal
B}({\cal H}^{\otimes n})$. Let the corresponding outputs for the
$i$-th branch of the channel be denoted by
$\sigma_{\alpha,i}^{(n)}$, i.e. \begin{equation}
\sigma_{\alpha,i}^{(n)} = \Phi_i^{\otimes n}(\rho_\alpha^{(n)}) =
\sigma^{i}_{\alpha,1} \otimes \dots \otimes \sigma_{\alpha,n}^i, \
\sigma_{\alpha,j}^i = \Phi_i(\rho_{\alpha,j}).
\end{equation}
Further define \begin{equation} {\bar \sigma}_i^{(n)} =
\frac{1}{|{\cal M}_n|} \sum_{\alpha \in {\cal M}_n}
\sigma_{\alpha,i}^{(n)} \end{equation} and \begin{equation} {\bar
\sigma}_{i,j} = \frac{1}{|{\cal M}_n|} \sum_{\alpha \in {\cal
M}_n} \sigma_{\alpha,j}^i\ . \end{equation} Let Bob's POVM
elements corresponding to the codewords $\rho_\alpha^{(n)}$ be
denoted by $E_\alpha^{(n)}$, $\alpha = 1,\dots,|{\cal M}_n|$. We
may assume that Alice's messages are produced uniformly at random
from the set ${\cal M}_n$. Then Bob's average probability of error
is given by
\begin{equation}
{\bar p}_e^{(n)} := 1- \frac{1}{|{\cal M}_n|} \sum_{\alpha \in
{\cal M}_n} \Trace\,\left[ \Phi^{(n)}(\rho_\alpha^{(n)})
E_\alpha^{(n)} \right].
\end{equation} We also define the average error corresponding
to the $i^{th}$ branch of the channel as
\begin{equation}
{\bar p}_{i,e}^{(n)} := 1- \frac{1}{|{\cal M}_n|} \sum_{\alpha \in
{\cal M}_n} \Trace\,\left[ \Phi_i^{\otimes n}(\rho_\alpha^{(n)})
E_\alpha^{(n)} \right].
\end{equation}
so that \begin{equation} {\bar p}_e^{(n)} = \sum_{i=1}^M \gamma_i
{\bar p}_{i,e}^{(n)}. \label{errpr}
\end{equation}

Let $X^{(n)}$ be a random variable with a uniform distribution
over the set ${\cal M}_n$, characterizing the classical message
sent by Alice to Bob. Let $Y_i^{(n)}$ be the random variable
corresponding to Bob's inference of Alice's message, when the
codeword is transmitted through the $i^{th}$ branch of the
channel. It is defined by the conditional probabilities
\begin{equation}
\PP\,[{Y_i^{(n)}} = \beta\,|\, X^{(n)} = \alpha] = \Trace\,
[\Phi_i^{\otimes n}(\rho_{\alpha}^{(n)}) E_{\beta}^{(n)}].
\end{equation} By Fano's inequality,
\begin{equation} h({\bar p}_{i,e}^{(n)}) + {\bar p}_{i,e}^{(n)}
\log(|{\cal M}_n|-1) \geq H(X^{(n)}\,|\, Y_i^{(n)}) = H(X^{(n)}) -
H(X^{(n)}\,:\, Y_i^{(n)}). \label{Fano}
\end{equation}
Here $h(\cdot)$ denotes the binary entropy and $H(\cdot)$ denotes
the Shannon entropy. Using the Holevo bound and the subadditivity
of the von Neumann entropy we have
\begin{eqnarray}
H(X^{(n)}\,:Y_i^{(n)}) &\leq& S\left(\frac{1}{|{\cal M}_n|}
\sum_{\alpha \in {\cal M}_n} \Phi_i^{\otimes n}
(\rho_{\alpha}^{(n)}) \right) - \frac{1}{|{\cal M}_n|}
\sum_{\alpha \in {\cal M}_n} S \left( \Phi_i^{\otimes n}
(\rho_{\alpha}^{(n)}) \right) \nonumber \\ &=& S \left(
\frac{1}{|{\cal M}_n|} \sum_{\alpha \in {\cal M}_n}
\sigma_{\alpha,i}^{(n)} \right) - \frac{1}{|{\cal M}_n|}
\sum_{\alpha \in {\cal M}_n} S( \sigma_{\alpha_i}^{(n)}) \non
\\ &\leq& \sum_{j=1}^n \left[ S\left({\bar \sigma}_{i,j} \right) -
\frac{1}{|{\cal M}_n|} \sum_{\alpha \in {\cal M}_n} S
\left( \sigma_{\alpha,j}^i \right) \right] \nonumber \\
&=&  \sum_{j=1}^n \chi_i\left( \left\{ \frac{1}{|{\cal M}_n|},
\rho_{\alpha,j} \right\}_{\alpha \in {\cal M}_n} \right) \non \\
&=& \sum_{j=1}^n \frac{1}{|{\cal M}_n|} \sum_{\alpha \in {\cal
M}_n} S \left( \sigma_{\alpha,j}^i\,||\, {\bar \sigma}_{i,j}
\right). \label{Holevosubadd}
\end{eqnarray}
The latter expression can be rewritten using Donald's identity:
\begin{equation} \sum_j p_j S(\omega_j\,||\,\rho) = \sum_j p_j
S(\omega_j\,||\,{\bar \omega}) + S({\bar \omega}\,||\, \rho),
\end{equation} where ${\bar \omega} = \sum_j p_j \omega_j$. We
apply this with $\rho$ replaced by \begin{equation} {\bar
\sigma}_i = \frac{1}{n |{\cal M}_n|} \sum_{j=1}^n \sum_{\alpha \in
{\cal M}_n} \sigma_{\alpha,j}^i \end{equation} and the sum
replaced by a double sum over $j$ and $\alpha$ with states
$\sigma_{\alpha,j}^i$. This yields
\begin{equation} \frac{1}{n |{\cal M}_n|} \sum_{j=1}^n \sum_{\alpha \in
{\cal M}_n} S(\sigma_{\alpha,j}^i \,||\, {\bar \sigma}_{i,j}) =
\frac{1}{n |{\cal M}_n|} \sum_{j=1}^n \sum_{\alpha \in {\cal M}_n}
S(\sigma_{\alpha,j}^i \,||\, {\bar \sigma}_i) + S({\bar \sigma}_i
\,||\, {\bar \sigma}_{i,j}). \end{equation} But, it follows from
convexity of the relative entropy that the second term on the
right-hand side is zero:
\begin{equation} 0 \leq S({\bar \sigma}_i
\,||\, {\bar \sigma}_{i,j}) \leq  \frac{1}{n} \sum_{j=1}^n S({\bar
\sigma}_{i,j} \,||\, {\bar \sigma}_{i,j}) = 0.
\end{equation}
Inserting into \reff{Holevosubadd} we now have: \begin{equation}
\frac{1}{n} H(X^{(n)}\,:\,Y_i^{(n)}) \leq \frac{1}{n |{\cal M}_n|}
\sum_{j=1}^n \sum_{\alpha \in {\cal M}_n} S(\sigma_{\alpha,j}^i
\,||\, {\bar \sigma}_i) = \chi_i \left( \left\{ \frac{1}{n |{\cal
M}_n|} , \rho_{\alpha,j} \right\}_{(\alpha,j)} \right).
\end{equation}

Fano's inequality  \reff{Fano} now yields \begin{equation} h({\bar
p}_{i,e}^{(n)}) + {\bar p}_{i,e}^{(n)} \log\,{|{\cal M}_n|} \geq
\log\,{|{\cal M}_n|} - n\,\chi_i \left( \left\{ \frac{1}{n |{\cal
M}_n|} , \rho_{\alpha,j} \right\}_{(\alpha,j)} \right),
\end{equation} However, since
\begin{equation} C(\Phi) \geq \bigwedge_{i=1}^M \chi_i \left(
\left\{ \frac{1}{n |{\cal M}_n|} , \rho_{\alpha,j}
\right\}_{(\alpha,j)} \right)
\end{equation} and $R = \frac{1}{n} \log |{\cal M}_n| > C(\Phi)$,
there must be at least one branch $i$ such that
\begin{equation} {\bar p}_{i,e}^{(n)} \geq 1 - \frac{C(\Phi)+
{1}/{n}}{R} > 0.\label{errpr2} \end{equation} We conclude from
(\ref{errpr}) and \reff{errpr2} that \begin{equation} {\bar
p}_e^{(n)} \geq  \left(1 - \frac{C(\Phi)+ {1}/{n}}{R}
\right)\,\bigwedge_{i=1}^M {\gamma_i}.
\end{equation} \qed

\section*{Acknowledgements} The authors would like to thank Andreas Winter
for a helpful suggestion. They are also grateful to 
Igor Bjelakovi\'c for carefully reading the paper and pointing
out some typos. This work was supported by the European Commission 
through the Integrated Project FET/QIPC "SCALA".

\end{document}